\documentclass[twocolumn,preprintnumbers,amsmath,amssymb]{revtex4}

\usepackage{graphicx}  
\usepackage{dcolumn}   
\usepackage{bm}        
\usepackage{amssymb}   
\usepackage{epsfig}
\usepackage{color}

\input colordvi
\epsfverbosetrue

\def\vr{\vec{\rm r}}

\begin{document}
\title{Quantum Monte Carlo facing the Hartree-Fock symmetry dilemma: The case of hydrogen rings}

\author{Peter Reinhardt$^{1)}$\footnote{Corresponding author:   email:
      {\tt
      Peter.Reinhardt@upmc.fr} ;  FAX: +33 1 44 27 41 17}, Julien Toulouse$^{1)}$,
      Roland Assaraf$^{1)}$, C. J. Umrigar$^{2)}$, Philip E. Hoggan$^{3)}$}
\affiliation{$^{1}$Laboratoire de Chimie Th\'eorique, Universit\'e Pierre et Marie Curie and CNRS, 4 place Jussieu, 75252 Paris, France}
\affiliation{$^{2}$Laboratory of Atomic and Solid State Physics, Cornell University, Ithaca, New York 14853, USA}
\affiliation{$^{3}$LASMEA, Universit\'e Blaise Pascal and CNRS,  24 avenue des Landais, 63177 Aubi\`ere, France}

\date{\today}

\begin{abstract}
When using Hartree-Fock (HF) trial wave functions in quantum Monte Carlo calculations, one faces, in case of HF instabilities, the HF symmetry dilemma in choosing between the symmetry-adapted solution of higher HF energy and symmetry-broken solutions of lower HF energies. In this work, we have examined the HF symmetry dilemma in hydrogen rings which present singlet instabilities for sufficiently large rings. We have found that the symmetry-adapted HF wave function gives a lower energy both in variational Monte Carlo and in fixed-node diffusion Monte Carlo. This indicates that the symmetry-adapted wave function has more accurate nodes than the symmetry-broken wave functions, and thus suggests that spatial symmetry is an important criterion for selecting good trial wave functions. 
\end{abstract}

\maketitle

\newpage

\section{Introduction}

It is well known that the symmetry-adapted solution of the nonlinear Hartree-Fock (HF) equations of an electronic system is sometimes unstable. An unstable solution corresponds to a saddle point of the energy as a function of the orbital parameters, and breaking of space and/or spin symmetries of the wave function then necessarily leads to one or several lower-energy HF solutions. The stability conditions of the HF equations were first formulated by Thouless~\cite{Tho-BOOK-61}, and the different instabilities were first categorized by {\v C}i{\v z}ek and Paldus~\cite{CizPal-JCP-67,PalCiz-JCP-70,CizPal-JCP-70,PalCis-JCP-71,CizPal-PRA-71,PalCiz-PRA-70}. For closed-shell systems, one may encounter ``singlet instabilities'' when only spatial symmetry is broken, and ``triplet (or nonsinglet) instabilities'' when spin symmetry is also broken. There is thus a symmetry dilemma~\cite{Low-RMP-63} in choosing between the symmetry-adapted wave function of higher HF energy and a symmetry-broken wave function of lower HF energy, in particular as a reference for a post-Hartree-Fock calculation.

A clear example is provided by closed-shell hydrogen rings H$_{4n+2}$ with equal bond lengths~\cite{ReiMal-JCP-99} (see, also, Ref.~\onlinecite{BenPal-JCP-80}). The symmetry-adapted HF solution exhibits singlet instabilities for sufficiently large numbers of hydrogen atoms, and one can obtain symmetry-broken HF solutions with orbitals localizing on either the atoms or the bonds. However, both M{\o}ller-Plesset perturbation theory and linearized coupled cluster doubles theory (also called CEPA--0 or DMBPT--$\infty$) give a lower total energy when starting from the symmetry-adapted solution than when starting from the symmetry-broken solutions, which casts doubts on the physical significance of the symmetry-broken solutions. Of course, the symmetry dilemma would be removed with a full configuration-interaction calculation which must give one unique solution, independent of the orbitals used.

Quantum Monte Carlo (QMC) approaches are alternatives to the traditional quantum chemistry methods~\cite{HamLesRey-BOOK-94,NigUmr-BOOK-99,FouMitNeeRaj-RMP-01}. The two most commonly used variants are variational Monte Carlo (VMC) which simply evaluates the energy of a flexible trial wave function by stochastic sampling, and fixed-node diffusion Monte Carlo (DMC) which improves upon VMC by projecting the trial wave function onto the ground state subject to the condition that the nodes of the projected wave function are the same as those of the trial wave function. For large systems, the most common form of the trial wave function is a Jastrow factor multiplied by a fixed HF determinant (though for small systems one can do much better by replacing the HF determinant by a linear combination of optimized Slater determinants). If a system exhibits HF instabilities, then QMC also faces the symmetry dilemma in choosing between different HF wave functions. Indeed, different HF wave functions necessarily lead to different energies not only in VMC, but also in DMC since the nodes of these HF wave functions are generally different. This symmetry dilemma in DMC is only due to the fixed-node approximation, since without this approximation DMC would give one unique solution, independent of the trial wave function. Of course, for systems that are not very large, symmetry breaking could probably be avoided in the first place by optimizing the orbitals within VMC~\cite{TouUmr-JCP-07,TouUmr-JCP-08}, instead of using fixed HF orbitals.

In this work, we study the impact of the HF symmetry dilemma for QMC in hydrogen rings H$_{4n+2}$. In Sec. II, we review the HF symmetry-breaking problem in these systems, and discuss the effect of using a Slater basis versus a Gaussian basis. In Sec. III, we explain the QMC methodology and report our VMC and DMC results. Our conclusions are summarized in Sec. IV.

\section{Hartree-Fock symmetry breaking}

\begin{figure*}
\includegraphics[scale=0.6,angle=0]{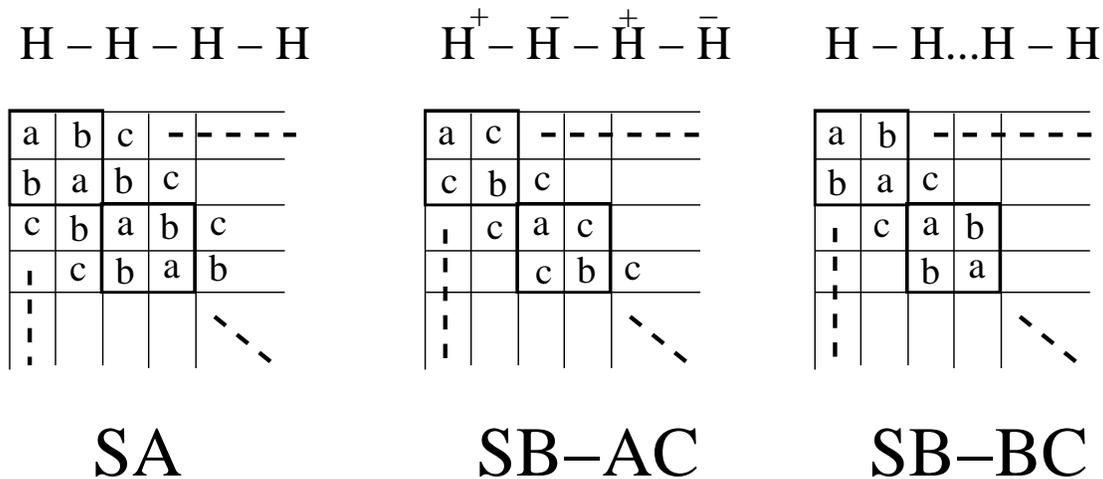}
\caption{Structure of the one-particle density matrix for the HF symmetry-adapted (SA) and symmetry-broken (SB-AC and SB-BC) solutions.}
\label{densmat}
\end{figure*}

In previous studies~\cite{ReiMal-JCP-98,ReiMal-JCP-99}, the electronic structure of periodic rings of $4n+2$ evenly spaced hydrogen atoms (with a fixed distance of $r_{\text{H}-\text{H}} = 0.74747$ \AA) has been investigated. The number of hydrogen atoms is restricted to $4n+2$ in order to obtain a possible closed-shell single-determinant solution with $2n+1$ occupied orbitals. The symmetry-adapted HF wave function has a metallic character (in the limit of an infinite ring) and can be expressed with either delocalized canonical orbitals or localized Wannier orbitals (which do not have an exponential decay). The canonical orbitals, except for the lowest one, are doubly degenerate, and in a minimal basis the orbital coefficients are fixed by the cyclic symmetry. Besides the symmetry-adapted (SA) solution, two different symmetry-broken HF solutions of lower energy can be obtained beyond critical ring sizes, when using unit cells of 2 hydrogen atoms. One solution corresponds to orbitals localized on hydrogen atoms and is referred to as the symmetry-broken atom-centered (SB-AC) solution, while the other corresponds to orbitals localized on bonds and is referred to as the symmetry-broken bond-centered (SB-BC) solution. The SB-BC solution has the lowest HF energy. The three solutions can be schematically described as $\cdots$H$\cdots$H$\cdots$, $\cdots$H$^+\cdots$H$^-\cdots$, and $\cdots$H---H$\cdots$. In each case, the symmetry breaking is accompanied by an opening of an energy gap between occupied and virtual orbitals, and orbitals decay much more rapidly than for the symmetry-adapted solution~\cite{ReiMal-JCP-99} , in agreement with the theoretical result of Kohn~\cite{Koh-CPL-93}.

In order to distinguish the three different wave functions, one may look at the one-particle density matrix $P$
\begin{equation}
  P_{\alpha\beta}\ = \ 2\,\sum_{i\in occ.} c_{\alpha\,i}\,c_{\beta\,i},
\end{equation}
containing the coefficients $c_{\alpha i}$ of the occupied molecular orbitals
$\varphi_i(\vr)$,
\begin{equation}
  \varphi_i(\vr)\ = \ \sum_\alpha\,c_{\alpha\,i}\ \chi_\alpha(\vr),
\end{equation}
expanded in a minimal set of atom-centered basis functions, i.e.\ one single basis function  $\chi_\alpha(\vr)$ per hydrogen atom. As depicted in Figure~\ref{densmat}, for the SA solution, we see equal elements on the diagonal and the sub-diagonals of the density matrix. For the SB-AC solution, an alternation of element values on the diagonal of the density matrix is obtained, but equal elements on the first sub-diagonal, and for the SB-BC solution we have equality of the diagonal elements and alternation on the first sub-diagonal.

In Ref.~\onlinecite{ReiMal-JCP-99}, a minimal Gaussian basis set (five $s$ Gaussian functions contracted to one single basis function for each hydrogen atom) was used. However, Gaussian basis functions are not appropriate for all-electron QMC calculations. They give large statistical fluctuations due to their incorrect vanishing gradient at the nuclear positions. It is thus much preferable to use Slater basis functions which correctly have a non-zero gradient on the nuclei and an exponential decay at large distance. In this work, we use a minimal Slater basis set (one $1s$ Slater function on each hydrogen atom) with an exponent of 1.17, which is smaller than the optimal exponent of 1.24 for an isolated H$_2$ molecule. Spin-restricted HF (and MP2) calculations were performed with an experimental code for ring systems, employed already for the previous studies~\cite{Ort-PROG-XX}. The necessary integrals over Slater functions have been calculated with the program SMILES~\cite{FerLopAguEmaRam-IJQC-01}. In order to obtain the symmetry-broken HF solutions, we start from a set of localized Wannier orbitals describing either an ionic situation or an explicit bond in the two-atom unit cell, and use an iterative configuration interaction procedure using singly excited determinants~\cite{Dau-CPL-74,ReiMalPovRub-IJQC-97} instead of diagonalizing a Fock operator.

\begin{table*}
  \caption{HF energy differences of H$_{4n+2}$ rings, per H$_2$ cell in mhartree, between the symmetry-broken solutions (SB-AC and SB-BC) and the
    symmetry-adapted (SA) one, for the Gaussian and Slater basis sets.}
  \begin{ruledtabular}
    \begin{tabular}{lcccc}
           & \multicolumn{2}{c}{Gaussian basis} &  \multicolumn{2}{c}{Slater basis}  \\
            \cline{2-3} \cline{4-5}
      4$n$+2 & $E$(SB-AC)-$E$(SA) & $E$(SB-BC)-$E$(SA) & $E$(SB-AC)-$E$(SA) & $E$(SB-BC)-$E$(SA) \\
      \colrule
      42  &    ---        &    ---       &   ---       &   ---        \\
      46  &    ---        &  $-0.00258$  &   ---       &   --- \\
      50  &    ---        &  $-0.03555$  &   ---       &  $-0.00199$ \\
      54  &  $-0.00115$   &  $-0.08933$  &   ---       &  $-0.02739$ \\
      58  &  $-0.01900$   &  $-0.15116$  &   ---       &  $-0.06961$ \\
      62  &  $-0.04980$   &  $-0.21440$  &  $-0.00080$ &  $-0.11914$ \\
      66  &  $-0.08678$   &  $-0.27565$  &  $-0.01223$ &  $-0.17074$
    \end{tabular}
  \end{ruledtabular}
  \label{tab:Slater}
\end{table*}

Table~\ref{tab:Slater} reports the HF energy differences between the symmetry-broken solutions and the symmetry-adapted one for the Gaussian basis set of Ref.~\onlinecite{ReiMal-JCP-99} and the Slater basis set of the present study. With the Gaussian basis set, the departure of the SB-BC and SB-AC solutions from the SA solution occurs for H$_{46}$ and H$_{54}$ rings, respectively. With the Slater basis set, the onset of symmetry breaking takes place for larger rings, i.e. for H$_{50}$ and H$_{62}$ for SB-BC and SB-AC, respectively. In addition, for a fixed ring size, the lowering in energy of the symmetry-broken solutions is smaller with the Slater basis set. This is an indication that the Slater basis is better than the Gaussian basis, since the amount of symmetry breaking is usually larger for poorer wave functions. The HF total energies are indeed lower with the Slater basis, for example for H$_{42}$, the (SA) energy per two-atom cell is $-0.950252$ hartree with the Gaussian basis and $-0.997003$ hartree with the Slater basis.  As an additional verification of the usefulness of Slater functions, we can look at Kato's cusp condition~\cite{Kat-CPAM-57} at the nuclear positions:
\begin{equation}
\frac{1}{2}\,\frac{|\vec{\nabla}\rho(\vec{r}_n)|_{av.}}{\rho(\vec{r}_n)} \ = \ Z,
\end{equation}
where $\rho(\vec{r}_n)$ and $|\vec{\nabla}\rho(\vec{r}_n)|_{av.}$ are the density and the spherical average of density gradient at the nuclear positions, and $Z$ is the nuclear charge. For example, for the H$_{86}$ ring, we find 1.009 for SA, 1.012 for SB-BC, and 1.034 and 0.980 for SB-AC, close to the ideal value of $Z=1$. Possibly, symmetry breaking can be further reduced by using a larger Slater basis.

\section{Quantum Monte Carlo study}

QMC methods are sometimes believed to produce benchmarks in quantum chemistry, approaching the electronic correlation problem differently than common wave-function-based methods or density-functional theory. As QMC methods often rely on a HF trial wave function, it is interesting to check their sensitivity to HF symmetry breaking. We start by giving a brief overview of the VMC and DMC methods employed in this work.

\subsection{Brief overview of VMC and DMC}

We consider Jastrow-Slater trial wave function of the form
\begin{equation}
\Psi_T({\bf R})\ = \ J({\bf R}) \ \Phi({\bf R}),
\end{equation}
where $\bf R$ designates the electron coordinates, $\Phi({\bf R})$ is a HF determinant and $J({\bf R}) = e^{f(\{r_{ij},r_{Ii}\})}$ is a Jastrow correlation factor depending explicitly on the electron-electron distances $r_{ij}$ and the nucleus-electron distances $r_{Ii}$. In VMC, one calculates the energy as the expectation value of the Hamiltonian $H$ over the wave function $\Psi_T$ by stochastic sampling
\begin{eqnarray}
  E_\text{VMC}  & = & \int \Psi_T({\bf R}) H \Psi_T({\bf R}) d{\bf R}
  = \int \Psi_T({\bf R})^2 E_L({\bf R}) d{\bf R} \cr
  & \approx & \frac{1}{M}\sum_{i=1}^{M} E_L({\bf R}_i),
\end{eqnarray}
where $E_L({\bf R}) = H \Psi_T({\bf R}) / \Psi_T({\bf R})$ is the local energy, and the $M$ points ${\bf R}_i$ are sampled from $\Psi_T({\bf R})^2$ by a Metropolis algorithm. In DMC, one improves over the distribution $\Psi_T({\bf R})^2$ by generating another distribution $f({\bf R},\tau)$ obtained by evolving the importance-sampling Schr\"odinger equation in imaginary time $\tau = i\,t$
 \begin{eqnarray}
 -\frac{\partial}{\partial \tau}f({\bf R},\tau) =
  -\frac{1}{2} \vec{\nabla}^2 f({\bf R},\tau) \;\;\;\;\;\;\;\;\;\;\;\;\;\;\;\;\;\;
\nonumber\\
      + \vec{\nabla}
    \cdot\left( f({\bf R},\tau) 
      \frac{\vec{\nabla}\Psi_T({\bf R})}{\Psi_T({\bf R})}\right)
  + \left(E_L({\bf R}) - E_T\right) f({\bf R},\tau),
\end{eqnarray}
which resembles an ordinary diffusion equation with diffusion, drift and source terms on the right-hand side. This diffusion process is simulated stochastically with a population of walkers representing the distribution $f({\bf R},\tau)$. The trial energy $E_T$ is adjusted in the course of the calculation in order to maintain a stable population of walkers. After some iterations, the stationary distribution is obtained $f({\bf R},\tau\to\infty) = \Psi_\text{FN}({\bf R}) \Psi_T({\bf R})$ where $\Psi_\text{FN}({\bf R})$ is the fixed-node (FN) wave function, i.e. the best approximation to the ground-state wave function having the same nodes than the trial wave function. In practice, this fixed-node approximation is automatically enforced by using $\Psi_\text{FN}({\bf R}) \Psi_T({\bf R})$ as a positive probability density, meaning that $\Psi_\text{FN}({\bf R})$ must necessarily be of the same sign as $\Psi_T({\bf R})$. The DMC energy is then calculated as the statistical average of the local energy of the trial wave function over the mixed distribution $\Psi_\text{FN}({\bf R}) \Psi_T({\bf R})$.

The nodes of the wave function are the locations of the points ${\bf R}$ where the wave function vanishes. For a system of $N$ electrons in 3 dimensions, they form $(3N-1)$--dimensional hypersurfaces. A subset of these nodes is given by the antisymmetry property of the fermionic wave function with respect to the exchange of two electrons, which implies that the wave function vanishes when two same-spin electrons are at the same point in space. However, these ``Pauli'' (or exchange) nodes form only $(3N-3)$--dimensional hypersurfaces, and are therefore far from sufficient to determine the full nodal hypersurfaces (see, e.g., Ref.~\onlinecite{BreCepRey-INC-01, BreMorTar-JCP-05, Mit-PRL-06} for examples of simple atomic systems). Likewise, spatial symmetry is generally far from sufficient to specify the nodes. For a given system, different HF wave functions (of different spatial symmetries) share the same Pauli nodes, but otherwise generally have very different nodal hypersurfaces, and thus lead to different fixed-node errors in DMC energies.

\subsection{Computational details}

The QMC calculations have been performed with the program CHAMP~\cite{Cha-PROG-XX} on a massively parallel IBM BlueGene architecture using up to 4096 processors. The trial wave functions are constructed by multiplying the previously obtained HF wave functions by a Jastrow factor consisting of the exponential of the sum of electron-nucleus, electron-electron and possibly electron-electron-nucleus terms, written as systematic polynomial and Pad\'e expansions~\cite{Umr-UNP-XX} (see also Refs.~\onlinecite{FilUmr-JCP-96,GucSanUmrJai-PRB-05}). Some Jastrow parameters are fixed by imposing the electron-electron cusp condition, and the others are optimized with the linear energy minimization method in VMC~\cite{TouUmr-JCP-07,UmrTouFilSorHen-PRL-07,TouUmr-JCP-08}, using an accelerated Metropolis algorithm~\cite{Umr-PRL-93,Umr-INC-99}. The orbital and basis exponent parameters are kept fixed in this work. Once the trial wave functions have been optimized, we perform DMC calculations within the short-time and fixed-node approximations (see, e.g., Refs.~\onlinecite{GriSto-JCP-71,And-JCP-75,And-JCP-76,ReyCepAldLes-JCP-82,MosSchLeeKal-JCP-82}). We use an imaginary time step of $\Delta \tau = 0.01$ hartree$^{-1}$ in an efficient DMC algorithm having very small time-step errors~\cite{UmrNigRun-JCP-93}. We use a target population of 100 walkers per processor, and estimate statistical uncertainties with blocks of 1000 steps, which is larger than the energy autocorrelation time of about 50 steps. The statistical uncertainty of the energy per H$_2$ cell is smaller than $2\times 10^{-5}$ hartree.

The computational cost of optimizing the two-body and three-body terms in the Jastrow factor scales as the third power of the number of hydrogen atoms.  When restricting the Jastrow factor to the two-body terms only, the computational cost scales quadratically with the number of hydrogen atoms, indicating that it is the evaluation of the Jastrow factor that dominates the computational cost and not the evaluation of the Slater determinant. The large reduction of computational cost achieved by removing the three-body terms comes without a big loss in the VMC energy, and in principle no loss at all in the DMC energy. For example, for the H$_{26}$ ring system, we find a VMC energy of $-13.8894 \pm  0.0005$ hartree with the three-body term, and $-13.8430 \pm 0.0005$ hartree without the three-body term. The computational effort is about 20 times more time consuming in the former case. We thus use a two-body Jastrow only.  Another alternative, not employed in this work, is to use a short-range three-body Jastrow.

\begin{figure}
\includegraphics[scale=1.2,angle=0]{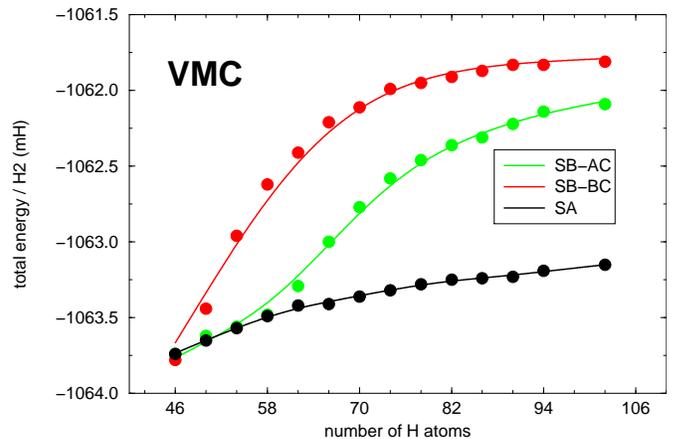}
\caption{VMC total energies of H$_{4n+2}$ rings, per H$_2$ cell in mhartree, for the symmetry-adapted (SA) and the two symmetry-broken (SB-AC and SB-BC) HF solutions from 46 to 102 hydrogen atoms. The statistical uncertainty is about the size of the point.}
\label{VMC_last}
\end{figure}
\begin{figure}
\includegraphics[scale=0.98,angle=0]{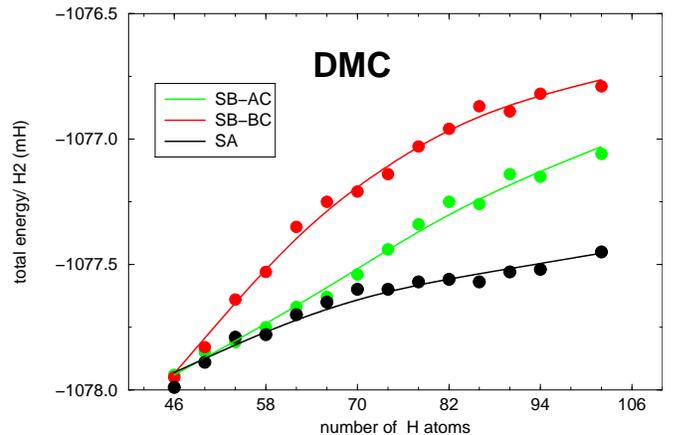}
\caption{DMC total energies of H$_{4n+2}$ rings, per H$_2$ cell in mhartree, for the symmetry-adapted (SA) and the two symmetry-broken (SB-AC and SB-BC) HF solutions from 46 to 102 hydrogen atoms. The statistical uncertainty is about the size of the point.}
\label{DMC}
\end{figure}

As the variance of the local energy of a ring of $n$ H$_2$ molecules is approximately $n$ times the variance for one H$_2$ molecule, the statistical uncertainty of the energy grows with the square root of $n$
\begin{equation}
\sigma \left( E(\text{H}_{2n}) \right) \sim \sqrt{\frac{V\left(E_L({\rm H}_{2n})\right)}{M}} \sim \sqrt{\frac{n \, V\left(E_L({\rm H}_2)\right)}{M}},
\end{equation}
where $M$ is the number of Monte Carlo iterations. Therefore, the statistical uncertainty of the energy per H$_2$ cell decreases as $1/\sqrt{n}$, and thus calculations aiming at a given statistical uncertainty for this quantity requite fewer steps for increasing ring sizes.

\subsection{Results}

Figure~\ref{VMC_last} shows the VMC energies per H$_2$ cell of the hydrogen rings with 46 to 102 atoms for the three HF solutions. As in the case of M{\o}ller-Plesset perturbation theory and linearized coupled cluster doubles theory, the energy ordering of the three solutions is reversed in comparison to HF, the SA wave function giving now the lowest total energy and the SB-BC solution giving the highest energy.  For H$_{86}$, we show in Table~\ref{tab:HFMP2VMCDMC} the QMC total energies and energy differences of the symmetry-adapted and symmetry-broken solutions. For comparison, we also report MP2 energies calculated with the same Slater basis set. The VMC total energy per H$_2$ cell lie about 50 mhartree below the MP2 energies, and the energy splittings between the different solutions are smaller than those in MP2, which shows the Jastrow factor does a good job of describing electron correlation.

\begin{table}[t]
  \caption{Total energy and energy differences, per H$_2$ cell in mhartree, of the symmetry-adapted (SA) and the symmetry-broken (SB-AC and SB-BC) HF solutions, for the H$_{86}$ ring, with the Slater basis set.}
  \begin{ruledtabular}
    \begin{tabular}{lccccc}
      method & $E$(SA) &   $E$(SB-AC)-$E$(SA) &  $E$(SB-BC)-$E$(SA)  \\
      \colrule
      HF   & $-996.15$   &  $-$0.14     & $-$0.39  \\
      MP2  & $-1016.22$   &  1.37        & 1.55     \\
      VMC  & $-1063.24$   &  0.93        & 1.36   \\
      DMC  & $-1077.57$   &  0.31        & 0.70    \\
    \end{tabular}
  \end{ruledtabular}
  \label{tab:HFMP2VMCDMC}
\end{table}

Figure~\ref{DMC} shows the corresponding DMC results. The energy ordering is the same as in VMC and MP2, the SA wave function giving the lowest DMC total energy, and thus the smallest fixed-node error. As shown in Table~\ref{tab:HFMP2VMCDMC}, the energy splittings between the different solutions are much smaller in DMC. This indicates that DMC is less sensitive to symmetry breaking than other correlation methods. It is an interesting feature for cases where symmetry breaking cannot be avoided. Of course, symmetry breaking would probably be avoided if the orbitals were reoptimized in the presence of the Jastrow factor within VMC, but this is computationally expensive for large systems.

\section{Conclusion}

When HF trial wave functions are used in QMC calculations, in case of HF instabilities QMC faces the HF symmetry dilemma in choosing between the symmetry-adapted solution of higher HF energy and symmetry-broken solutions of lower HF energies. In this work, we have examined the HF symmetry dilemma in hydrogen rings H$_{4n+2}$ which present HF singlet instabilities for sufficiently large ring sizes. We have shown that using a Slater basis set, instead of a Gaussian basis set, delays the onset of HF symmetry breaking until larger rings and slightly reduces the energy splittings between the symmetry-adapted and symmetry-broken wave functions. When using these different HF wave functions in VMC and DMC, we have found that the energy ordering is reversed; the symmetry-adapted wave function always giving the lowest energy. This confirms previous post-Hartree-Fock studies in showing that these symmetry-broken solutions are bad starting wave functions for correlated calculations. The fact that the symmetry-adapted wave function gives the lowest DMC energy indicates that this wave function has more accurate nodes than the symmetry-broken wave functions. The present experience thus suggests that spatial symmetry is an important criterion for selecting good trial wave functions. For systems that are not very large, the symmetry-breaking problem could probably be avoided altogether by optimizing the orbitals within the quantum Monte Carlo calculation, rather than using fixed HF orbitals.

\section*{Acknowledgment}
This work has been financed mainly through the DEISA network, project STOP-Qalm. All QMC calculations have been performed on the IBM Bluegene machines in J\"ulich and Munich (Germany). C.J.U. was supported in part by the NSF (Grant Nos. DMR-0908653 and CHE-1004603). The authors thank the staff of IDRIS (Orsay, France) for technical assistance to install, test and run the QMC program CHAMP on these machines. We also acknowledge using the Slater integral code SMILES (Madrid, Spain) for obtaining the HF starting wave functions. Discussions with P. Gori-Giorgi (Amsterdam, Netherlands) and J.-P. Malrieu (Toulouse, France) were very helpful for the project.


\end{document}